\documentstyle{mn}
\begin{document}
\def\simlt{\mathrel{\rlap{\lower 3pt\hbox{$\sim$}} \raise
        2.0pt\hbox{$<$}}} \def\simgt{\mathrel{\rlap{\lower
        3pt\hbox{$\sim$}} \raise 2.0pt\hbox{$>$}}} 
\title[Theoretical predictions on the clustering of SCUBA galaxies] 
{Theoretical predictions on the clustering of SCUBA galaxies and
implications for small-scale fluctuations at sub-mm wavelengths} 
\author[M. Magliocchetti, L. Moscardini,
        P. Panuzzo, G.L. Granato, G. De Zotti, L. Danese]
{M.Magliocchetti$^{1}$, L.Moscardini$^{2}$, P.Panuzzo$^{1}$,
G.L.Granato$^{3}$, G. De Zotti$^{3}$, L.Danese$^{1}$\\ $^{1}$
SISSA, Via Beirut 4, 34100, Trieste, Italy\\ $^{2}$
Dipartimento di Astronomia, Universit\`a di Padova, Vicolo
dell'Osservatorio 5, I-35122 Padova, Italy\\ $^{3}$
Osservatorio Astronomico, Vicolo dell'Osservatorio 5, I-35122
Padova, Italy\\} \maketitle
\vspace {7cm }
 
\begin{abstract} 
This paper investigates the clustering properties of SCUBA-selected
galaxies in the framework of a unifying scheme relating the formation
of QSOs and spheroids (Granato et al. 2000). The theoretical angular
correlation function is derived for different bias functions,
corresponding to different values of the ratio $M_{\rm halo}/M_{\rm
sph}$ between the mass of the dark halo and the final mass in stars.
SCUBA sources are predicted to be strongly clustered, with a
clustering strength increasing with mass. We show that the model
accounts for the clustering of Lyman-break Galaxies, seen as the
optical counterpart of low- to intermediate-mass primeval spheroidal
galaxies and is also consistent with the observed angular correlation
function of Extremely Red Objects. Best agreement is obtained for
$M_{\rm halo}/M_{\rm sph}=100$. We also consider the implications for
small scale fluctuations observed at sub-mm wavelengths by current or
forthcoming experiments aimed at mapping the Cosmic Microwave
Background (CMB). The predicted amplitude of the clustering signal in
the 350 GHz channel of the Planck mission strongly depends on the
halo-to-bulge mass ratio and may be of comparable amplitude to primary
CMB anisotropies for multipole numbers $l\simgt 50$.
\end{abstract}
\section{Introduction}
The advent of the 450-850 $\mu$m Submillimetre Common User Bolometric
Array (SCUBA) camera (Holland et al. 1999) has opened the possibility
of exploiting the enormous potential of sub-mm astronomy for
investigating the early phases of galaxy evolution, taking advantage
of the strongly negative K-correction which boosts the visibility of
high-$z$ galaxies (Smail, Ivison \& Blain 1997; Barger et al. 1998;
Hughes et al. 1998; Barger, Cowie \& Sanders 1999; Eales et al. 1999;
Blain et al. 1999a).

A variety of models has been proposed to reproduce the properties of
the submillimetre population (e.g. Guiderdoni et al. 1998; Blain et
al.  1999b; Devriendt, Guiderdoni \& Sadat 1999; Devriendt \&
Guiderdoni 2000; Takeuchi et al. 2000; Pearson 2000), in particular
the dramatic star-formation rate inferred from the data.

Granato et al. (2000 - hereafter Paper I) propose a unified scheme for
the formation of the QSO and spheroidal population. According to their
model, at the time of the onset of the QSO activity, the host galaxy
has already undergone a major star-formation event, with a
characteristic time $T_B$ ranging from $\sim 0.5$ to $\sim 2$~Gyr when
going from more to less massive hosts. The QSO onset practically stops
the star-formation process, and when eventually the reservoir of gas
fueling the Active Galactic Nuclei is exhausted, the host galaxy will
evolve in a passive way, with a spectrum becoming quickly red.  This
approach identifies SCUBA galaxies as the progenitors of QSO hosts and
provides an excellent description of (amongst others) the observed 850
$\mu$m number counts.

In this paper we present predictions on clustering properties of SCUBA
galaxies in the framework of the Granato et al. (2000) model. Such
properties are determined by the combination of the evolution of the
underlying mass fluctuations and the bias relating galaxy
overdensities to mass.  Since the bias factor is determined by
properties related to the process of galaxy formation, it follows that
measurements of clustering can lead to very precious insights on the
galaxy population producing the signal.

Unfortunately the number of SCUBA galaxies observed is still too small
to allow for any clustering measurement. The model by Granato et al.
(2000) however establishes a tight link between this population and
Lyman-break galaxies (Steidel et al. 1996; LBGs hereafter),
interpreted as lower-mass/lower-submm luminosity sources. Such model
can thus be tested against measurements of LBG clustering (Giavalisco
et al. 1998). A further test is provided by the recent detection
(Daddi et al. 2000) of strong clustering of Extremely Red Objects
(EROs), which were found to be the optical counterparts of two of the
still few SCUBA sources having reliable optical identifications (Smail
et al. 1999).

An important issue related to the above discussion is the effect of
clustering on small scale anisotropies at mm/sub-mm wavelengths, where
experiments to map the Cosmic Microwave Background either have been
(e.g. BOOMERanG, MAXIMA, TopHat, ACBAR, ARCHEOPS) or are about to be
carried out (notably the ESA's Planck mission).  Small scale
fluctuations of the background intensity due to clustering of
unresolved sources are proportional to the amplitude of their
two-point correlation function and to their volume emissivity (see,
e.g., De Zotti et al.  1996). The deep SCUBA surveys have demonstrated
that the bulk of the extragalactic background at $850\,\mu$m comes
from a relatively small number of sources (at least ten time less than
those accounting for the bulk of the optical/UV background) at
substantial redshifts. Combining the (rather limited) direct
spectroscopic information (e.g. Frayer 2000) with estimates based on
the sub-mm to radio spectral index (Carilli \& Yun 1999), it is
concluded that the median redshift of the sub-mm population brighter
than $\sim 1\,$mJy is probably $\sim 2.5$--3 (Smail et al. 2000). It
then follows that most of the background comes from a population of
extremely luminous (hence presumably very massive) high-$z$ galaxies
which probably correspond to the rare, high density peaks in the
primordial matter distribution.  The standard structure formation scenario 
predicts that these peaks are strongly clustered (Kaiser 1984; 
Baron \& White 1987).  Also, according to the currently
favoured view, SCUBA sources probably correspond to the phase of
intense star formation in large spheroidal galaxies, a process that
appears to be completed for $z >1$; as a consequence, the relevant
redshift range is limited and dilution of the clustering signal is
relatively small.

The amplitude of intensity fluctuations due to clustering is
correspondingly expected to be large. Scott \& White (1999) worked out
a preliminary estimate adopting for sub-mm sources the angular
correlation function determined for $z\sim 3$ LBGs (which however are
mostly undetectable by SCUBA; see Chapman et al. 2000).  Here we
provide quantitative estimates of this effect using a technique
developed by Peacock \& Dodds (1996), Matarrese et al. (1997) and
Moscardini et al. (1998), which takes into account the effects of the
past light-cone, the linear and non-linear growth of clustering and
the evolution of the bias factor. The epoch-dependent mass
distribution of sources is modelled following Granato et al. (2000).

The plan of the paper is as follows. Section 2 will be devoted to the
analysis of the clustering of SCUBA galaxies and comparisons with the
observed clustering properties of LBGs and EROs, while in Section 3 we
use the findings of Section 2 to make predictions on the temperature
fluctuations produced by unresolved SCUBA-like sources and to discuss
their implications for experiments aimed at mapping the CMB at sub-mm
wavelengths. Section 4 summarizes our conclusions.

Throughout the paper we will assume a cosmological model with
$h_0=0.7$, $\Omega_0=0.3$, $\Lambda=0.7$, and adopt the normalization
of the rms mass fluctuation $\sigma_8=1$ determined by the 4-year COBE
data (Bunn \& White 1997).

\section{Clustering of SCUBA sources}
The angular two-point correlation function $w(\theta)$ is defined as
the excess probability, $\delta P$, with respect to a random Poisson
distribution, of finding two sources in the solid angles
$\delta\Omega_1$ $\delta\Omega_2$ separated by an angle $\theta$, and
it is defined by
\begin{eqnarray}
\delta P=n^2\delta\Omega_1\delta\Omega_2\left[1+w(\theta)\right]
\label{eqn:wthetadef}
\end{eqnarray}
where $n$ is the mean number density of objects in the particular
catalogue under consideration.

The theoretical expression for $w(\theta)$ can be derived from its
spatial counterpart $\xi$ by projection via the relativistic Limber
equation (Peebles 1980): ~~
\begin{eqnarray}
w(\theta)=2\:\frac{\int_0^{\infty}\!\!\int_0^{\infty}N^2(z)\;b_{\rm
eff}^2(M_{\rm min}, z)\;
(dz/dx)\;\xi(r,z)\;dz\;du}{\left[\int_0^{\infty}N(z)\;dz\right]^2},
\label{eq:limber}
\end {eqnarray}     
where $x$ is the comoving radial coordinate,
$r=(u^2+x^2\theta^2)^{1/2}$ (for a flat universe and in the small
angle approximation: $\theta \ll 1\,$rad), and $N(z)$ is the number of
objects within the shell ($z,z+dz$).

\begin{figure}
\vspace{8cm} \includegraphics{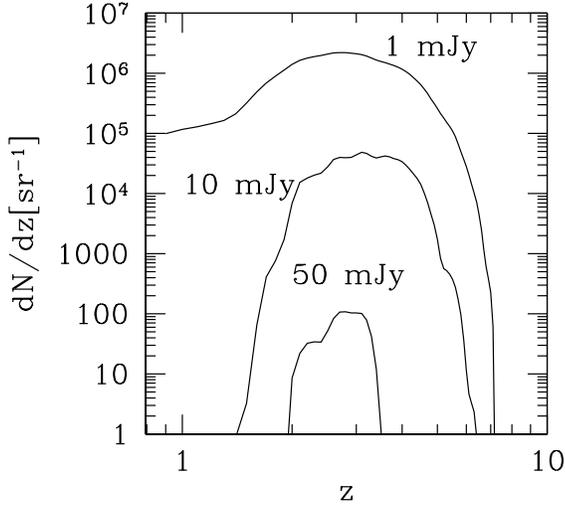}
\caption{Predicted redshift distribution for SCUBA galaxies
respectively brighter than 1, 10 and 50 ~mJy. The Figure is adapted
from Paper I.
\label{fig:N_z}}
\end{figure}

The mass-mass correlation function $\xi(r,z)$ to be inserted in
eq.(\ref{eq:limber}) has been obtained following the work by Peacock
\& Dodds (1996, but also see Magliocchetti et al., 2000), extended by
Matarrese et al. (1997) and Moscardini et al. 1998) to redshifts $z>
0$, which provides an analytical way to derive the trend of $\xi(r,z)$
both in the linear and non-linear regime. Note that $\xi(r,z)$ only
depends on the underlying cosmology and on the normalization of
$\sigma_8$.  The relevant properties of SCUBA galaxies are included in
the redshift distribution of sources $N(z)$, where $N(z)\;dz$ gives the 
number of objects in the shell $(z,z+dz)$ (given in Paper I and
shown in Fig.~\ref{fig:N_z} for fluxes respectively greater than 1,
10, 50~mJy), and in the bias factor $b_{\rm eff}(M_{\rm min},z)$.

The effective bias factor $b_{\rm eff}(M_{\rm min}, z)$ of all haloes
with masses greater than some threshold $M_{\rm min}$ is obtained by
integrating the quantity $b(M,z)$ - representing the bias of
individual haloes of mass $M$ - opportunely weighted by the number
density $n_{SCUBA}(M,z)$ of SCUBA sources:
\begin{eqnarray}
b_{\rm eff}(z)=
\frac{\int_{M_{\rm min}}^{\infty} dM\;b(M,z)\;n_{\rm SCUBA}(M,z)}
{\int_{M_{\rm min}}^{\infty} dM\;n_{\rm SCUBA}(M,z)} \ .
\label{eq:b}
\end{eqnarray}
Note that as $n_{\rm SCUBA}$ can be thought as the fraction of haloes
hosting a galaxy in the process of forming stars, its expression can
also be written as $n_{SCUBA}(M,z)=n(M,z)\;T_B/t_h$, where $n(M,z)\; dM$ is
the mass spectrum of haloes with masses between $M$ and $M+dM$ (Press
\& Schechter, 1974; Sheth \& Tormen, 1999), $T_B$ is the duration of
the star-formation burst and $t_h$ is the life-time of the haloes in
which these objects reside (see Martini \& Weinberg, 2000).  The
duration of the star-formation phase $T_B$ varies with the mass of the
spheroid undergoing the burst (see PaperI), assuming a value
$T_b=0.5$~Gyr in the case of massive objects (i.e. with bulge masses
$M_{sph}\simgt 1.5\; 10^{11} M_{\odot}$) and increasing when going to
lower masses.\\ We take the expression for $b(M,z)$ to be plugged into
eq.(\ref{eq:b}) from Jing (1998):
\begin{eqnarray}
b(M,z)=\left(1+\frac{1}{\delta_c}\left[\frac{\delta_c^2(z)}{\sigma^2(M)}-1
\right]\right)\!\!\!
\left(\frac{\sigma^4(M)}{2\;\delta_c^4(z)}+1\right)^{\!\!\!\!
\frac{(6-2n_{\rm eff})}{100}}
\label{eq:bj}
\end{eqnarray}
with $\delta_c(z)=\delta_{c}/D(z)$, $D(z)$ being the linear growing
factor and $\delta_c$ the critical density for collapse of a
homogeneous spherical perturbation ($\delta_c=1.686$ in an Einstein-de
Sitter universe). $\sigma^2(M)$ is the variance of density
fluctuations on the mass scale $M$, and $n_{\rm eff}=-3-6(d{\rm ln}
\sigma/d{\rm ln}M)$. Note that the assumption of a linear bias is 
justified by the redshift distribution of the sources, all appearing 
for $z\simgt 0.75$ (see e.g. Somerville et al., 2001).

According to Granato et al. (2000), sources showing up in the SCUBA
counts can be broadly divided into three categories: low-mass (masses
in the range $M_{\rm sph}\simeq 10^9-10^{10}M_\odot$, duration of the
star formation burst $T_B\sim 2$~Gyr, and typical fluxes $S\simlt
1$~mJy), intermediate-mass ($M_{\rm sph}\simeq
10^{10}$--$10^{11}M_\odot$ and $T_B\sim 1$~Gyr) and high-mass ($M_{\rm
sph}\simgt 10^{11}M_\odot$, $T_B\sim 0.5$~Gyr, dominating the counts
at fluxes $S\simgt 5-10$~mJy).  Note that by $M_{\rm sph}$ we denote
the mass in stars at completion of the star formation process.\\ In
order to evaluate the bias factor in eq.(\ref{eq:b}) we then consider
two extreme cases for the ratio between the mass in stars and the mass
of the host dark halo: $M_{\rm halo}/M_{\rm sph}=100$ and $M_{\rm
halo}/M_{\rm sph}=10$.  Note that $M_{\rm halo}/M_{\rm sph}=10$
roughly corresponds to the ratio $\Omega_0/\Omega_{\rm baryon}$
between total and baryon density, where we adopted for the latter
quantity the standard value from primordial nucleosynthesis; this
corresponds to having assumed all the baryons to be locked into stars and, as 
a consequence, gives a conservative lower limit to the quantity 
$M_{\rm halo}/M_{\rm sph}$. $M_{\rm halo}/M_{\rm sph}=100$ is instead related 
to $\Omega_0/\Omega_{\star}$, $\Omega_{\star}$ being the present mass
density in visible stars (Persic \& Salucci, 1992; Fukugita et al. 1998).  
The likely value should be $M_{\rm sph}/M_{\rm halo}=1-3$~\% (cf. Paper I).

\begin{figure}
\vspace{8cm} \includegraphics{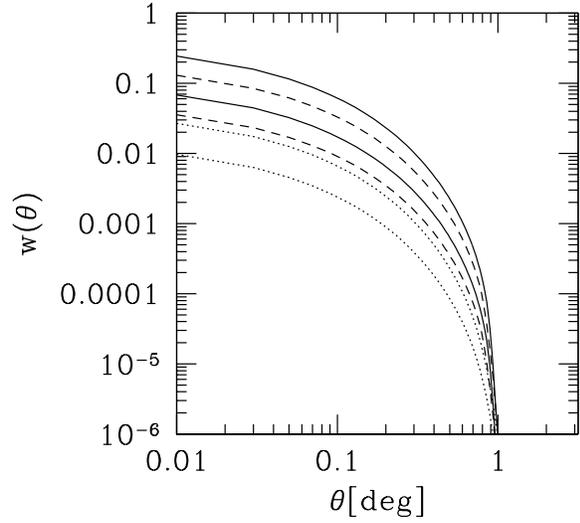}
\caption{Predictions for the angular correlation function $w(\theta)$
for different halo-to-bulge mass ratios and different flux cuts.
Solid lines are for sources brighter than 50~mJy, dashed lines for
sources brighter than 10~mJy, while dotted lines correspond to $S\ge
1$~mJy. Higher curves of each type correspond to $M_{\rm halo}/M_{\rm
sph}=100$, lower ones to $M_{\rm halo}/M_{\rm sph}=10$.
\label{fig:w_scuba}}
\end{figure}  

Armed with the above results we can finally evaluate the two-point
correlation function in eq.~(\ref{eq:limber}) for different $M_{\rm
halo}/M_{\rm sph}$ ratios and different flux cuts. Figure
\ref{fig:w_scuba} presents our predictions for $w(\theta)$,
respectively for a flux cut of 50 (solid line), 10 (dashed line) and 1
(dotted line) mJy. Higher curves of each kind correspond to the case
$M_{\rm halo}/M_{\rm sph}=100$, while lower curves refer to the
$M_{\rm halo}/M_{\rm sph}=10$ one.

The highest clustering amplitude is found for the brightest sources
($S \ge 50$~mJy). This is because they are associated to the most
massive dark halos and are therefore highly biased tracers of the dark
matter distribution. In addition, according to the model by Granato et
al. (2000), they have a rather narrow redshift distribution (see
Fig.~\ref{fig:N_z}), so that the dilution of the clustering signal is
minimum.\\ The very sharp drop of all the curves at $\theta\simeq
1^\circ$ is due to the absence of nearby ($z\simlt 1$, see
Fig.~\ref{fig:N_z}) objects. This reflects the notion that the
actively star-forming phase in spheroids is completed at $z> 1$.
 
The model introduced in PaperI establishes a strong link between SCUBA
sources and LBGs where this latter population is predicted to be the
low- to intermediate-mass tail of primeval spheroidal galaxies, with
$T_B \sim 1-2$~Gyr and a star-formation rate ranging from a few to a
hundred $M_\odot\; \hbox{yr}^{-1}$. \\
\begin{figure}
\vspace{8cm} \includegraphics{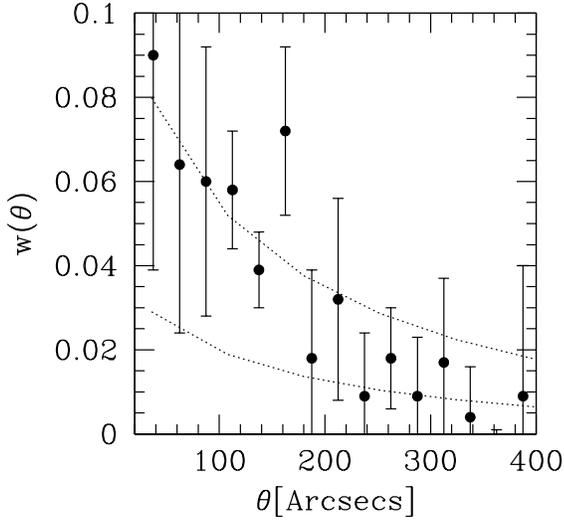}
\caption{Angular correlation function of LBGs. Model predictions are
shown by the dotted lines (the higher one is for $M_{\rm halo}/M_{\rm
sph}=100$, the lower for $M_{\rm halo}/M_{\rm sph}=10$), while data
points represent the Giavalisco et al. (1998) measurements.
\label{fig:w_LBG}}
\end{figure}  
Measurements of the angular correlation function of LBGs (Giavalisco
et al. 1998) then offer a way to test such model.  In
Fig.~\ref{fig:w_LBG} we plotted the expected $w(\theta)$ for sources
with $S\ge 1$~mJy (corresponding to $M_{\rm sph} \simgt 10^{10}
M_\odot$) within the redshift range $2.5 \le z \le 3.5$, covered by
the Giavalisco et al. (1998) sample. As in Fig.~\ref{fig:w_scuba}, the
higher curve is for $M_{\rm halo}/M_{\rm sph}=100$, the lower one for
$M_{\rm halo}/M_{\rm sph}=10$. The data points show the Giavalisco et
al.  (1998) measurements. Despite the large errors, it is clear that
the predicted trend for $w(\theta)$ correctly reproduces the data, a
better fit obtained for $M_{\rm halo}/M_{\rm sph}=100$. This result,
apart from providing a further proof of the tight link between SCUBA
galaxies and LBGs, also confirms the expectations of Paper I for a
small fraction (on the order of a few percent) of the total mass to be
confined into stars.

Strong clustering of EROs has been recently detected by Daddi et
al. (2000).  This result may again provide a relevant test for our
model since it seems likely for a substantial fraction of SCUBA
galaxies to be EROs with $R-K \geq 6$: at least two out of nine ($>
20\%$) sources in the Cluster Lens Survey belong to this class (Smail
et al. 1999), while other submillimeter surveys have discovered
additional EROs as counterparts of SCUBA sources (Ivison et
al. 2000). For their ERO sample with $K_s < 19.2$ and $R-K_s \geq
5.3$, Daddi et al. (2000) find $w(\theta) \simeq 0.07$ for $\theta
\simeq 0^\circ.1$, not far from the predictions in
Fig.~\ref{fig:w_scuba}.

\section{Small Scale Fluctuations from Discrete Sources}
\subsection{The Intensity Correlation Function}
An issue intimately connected with the analysis of galaxy clustering
performed in the previous Section is the study of the contribution of
unresolved sources (i.e.  sources with fluxes fainter than some
detection limit $S_d$) to the background intensity. Its general
expression is:
\begin{eqnarray}
I=\int_0^{S_d}
\frac{dN}{dS}\;S\;dS\;\;\;\;\;\;\;\;\;\;\;\;\;\;\;\;\;\;
\;\;\;\;\;\;\;\;\;\;\;\;\;\;\;\;\;\;\;\;\;\;\;\;\;\;\;\;\;\;\;\;\;\nonumber\\
=\frac{1}{4\pi}\int^{L_{\rm max}}_{L_{\rm min}}\!\!\!\!\!\!\!\!\!\!
d{\rm log}L\;L\;\int_{z(S_d,L)}^{z_{\rm max}}\!\!\!\!\!
dz\;\Phi(L,z)\;\frac{K(L,z)}{(1+z)^2}\;\frac{dx}{dz}
\label{eq:I}
\end{eqnarray}
(see e.g. De Zotti et al. 1996), where $dN/dS$ denotes the
differential number counts, $L_{\rm max}$ and $L_{\rm min}$ are
respectively the maximum and minimum local luminosity of the sources,
$K(L,z)=(1+z)L[\nu(1+z)]/L(\nu)$ is the K-correction, $z_{\rm max}$ is
the redshift when the sources begin to shine, $z(S_d,L)$ is the
redshift at which a source of luminosity $L$ is seen with a flux equal
to the detection limit $S_d$, $\Phi(L,z)$ is the luminosity function
(i.e. the comoving number density of sources per unit $d{\rm log}L$),
and $x$ is the comoving radial coordinate.

The intensity fluctuation $\delta I$ due to inhomogeneities in the
space distribution of unresolved sources is then given by
eq. (\ref{eq:I}), with the quantity $\Phi(L,z)$ replaced by $\delta
\Phi(L,z)$. It is easily shown that the angular correlation of such
intensity fluctuations
\begin{eqnarray}
C(\theta)=\langle \delta I(\theta^\prime, \phi^\prime)\; \delta
I(\theta'', \phi'')\rangle ,
\label{eq:ctheta}
\end{eqnarray} 
where $(\theta^\prime, \phi^\prime)$ and $(\theta'', \phi'')$ define
two positions on the sky separated by an angle $\theta$, can be
expressed as the sum of two terms $C_P$ and $C_C$, the first one due
to Poisson noise (i.e. fluctuations given by randomly distributed
objects), and the second one owing to source clustering.

Scott \& White (1999) have shown that if dusty galaxies cluster like
LBGs (Giavalisco et al. 1998), anisotropies due to clustering dominate
the Poisson ones at all angular scales. According to our model, LBGs
are the optical counterparts of the low- to intermediate- mass tail of
SCUBA galaxies; as illustrated in Figure~\ref{fig:w_scuba}, this makes
the clustering signal of LBGs a factor $\sim 10$ lower than the one
obtained for SCUBA galaxies; therefore we can safely assume the
Poisson term $C_P$ to be negligible in the following analysis and only
concentrate on temperature fluctuations caused by the $C_C$ term
(hereafter simply called $C$).

By making use of the quantities defined in Section 2 and of
eq.~(\ref{eq:I}), the clustering term in eq.~(\ref{eq:ctheta}) takes
the form:
\begin{eqnarray}
C(\theta)=\left({1\over 4\pi}\right)^2\!\!\!  \int_{z_{(L_{\rm
min},S_d})}^{z_{\rm max}}\!\!\!\!\!\!\!\!\!\!\!\!\!\!\!\!\!\!\!\!
\!\!\!\! dz\;b_{\rm eff}^2(z)\; \frac{j^2_{\rm eff}(z)}
{(1+z)^4}\left(\frac{dx}{dz}\right)^2
\int_0^\infty\!\!\!\!\!\!\!\!du\;\xi(r,z),
\label{eq:cth}
\end{eqnarray}
where
\begin{eqnarray}
j_{\rm eff}=\int_{L_{\rm min}}^{{\rm min}[L_{{\rm
max},L(S_d,z)}]}\!\!\!\!\!  \Phi(L,z)\; K(L,z)\; L\;d{\rm log}L
\label{eq:jeff}
\end{eqnarray}
is the effective volume emissivity (see Toffolatti et al. 1998).

\begin{figure*}
\vspace{8cm} \includegraphics{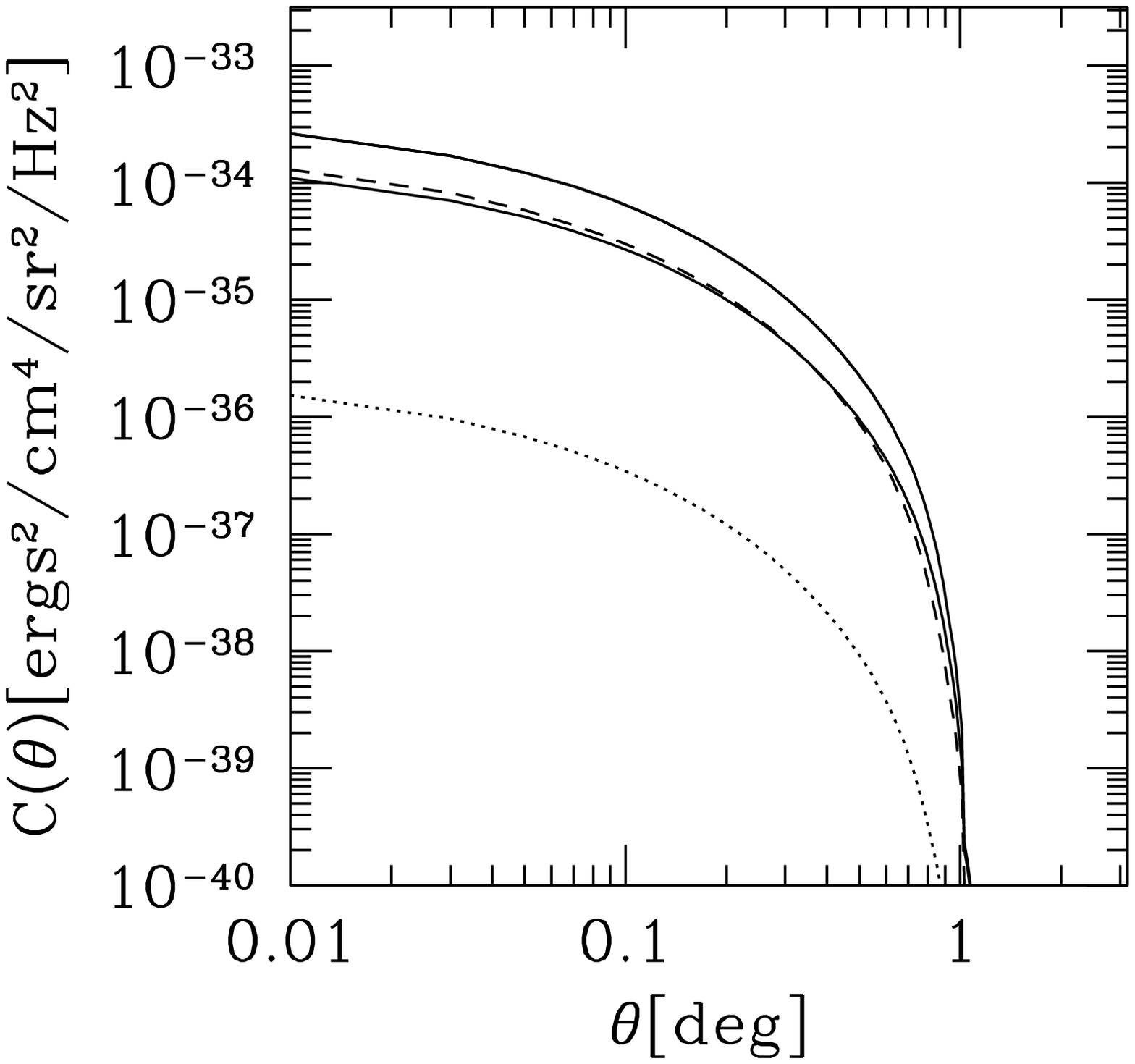} \includegraphics{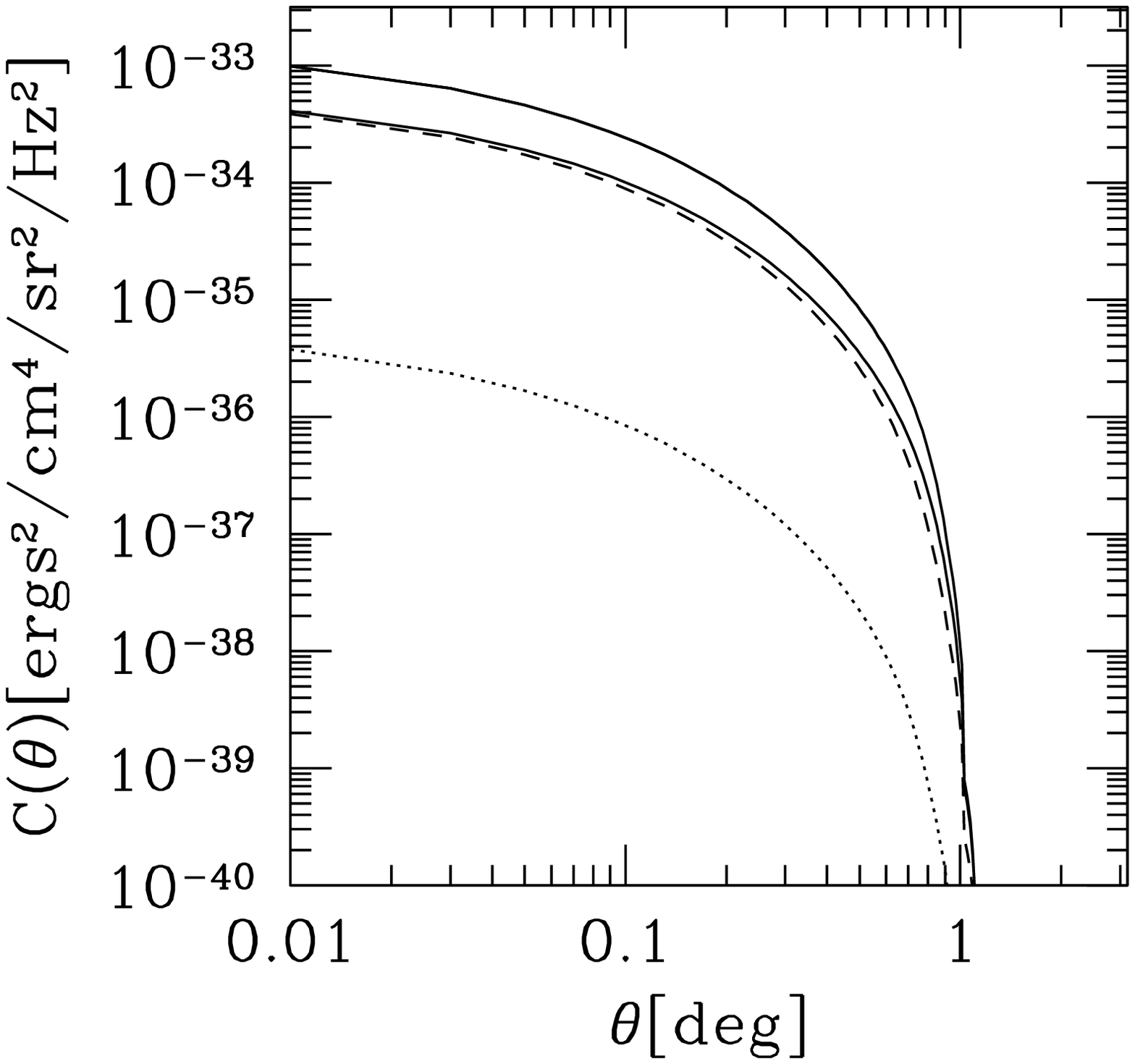}
\caption{Left panel: predictions for the correlation signal due to
intensity fluctuations of SCUBA galaxies in the case $M_{\rm
halo}/M_{\rm sph}=10$.  Dotted, dashed and solid lines are
respectively for spheroids of low ($M_{\rm sph}=10^9$--$10^{10}\;
M_{\odot}$), intermediate ($M_{\rm sph}=10^{10}$--$10^{11}\;
M_{\odot}$), and high ($M_{\rm sph} > 10^{11}\; M_{\odot}$) masses.
The two solid curves are obtained for different flux detection limits,
the 100~mJy case represented by the higher curve and the 10~mJy case
by the lower one.  Right panel: same as before but with a ratio
$M_{\rm halo}/M_{\rm sph}=100$.
\label{fig:lowb}}
\end{figure*}  

Following the prescriptions of Section 2, we have then evaluated
$C(\theta)$ in eq.(\ref{eq:cth}) separately for the three cases of
low-, intermediate- and high-mass objects, by plugging in
eq.~(\ref{eq:jeff}) the appropriate expressions for the luminosity
function.  Note that in this case, the quantity $b^2(M_{\rm min},z)$
should indeed be read as $b^2(M_{\rm min},M_{\rm max},z)$, where
$M_{\rm max}$ is the maximum halo mass corresponding to the maximum
visible bulge mass (i.e.  upper limit for the mass locked into stars
whose values are given in Section 2). This corresponds in
eq.(\ref{eq:b}) to a replacement in the upper limit of the integrals
of $\infty$ with $M_{\rm max}$.

Figure \ref{fig:lowb} shows predictions for the correlation signal due
to intensity fluctuations (in units ergs$^2$ cm$^{-4}$ Hz$^{-2}$
sr$^{-2}$) in the case of SCUBA galaxies.  Dotted, dashed and solid
lines are respectively for spheroids of $M_{sph}^{min}=10^9,\;10^{10}$
and $10^{11}\; M_{\odot}$.  In each plot the two solid curves
correspond to different detection limits, the 100~mJy case represented
by the higher curve and the 10~mJy case by the lower one.  Note that,
due to the steepness of the bright end of source counts, the results
are independent of $S_d$ if $S_d\ge 100$~mJy. Results are shown for
$M_{\rm halo}/M_{\rm sph}=10$ (left-hand panel) and $M_{\rm
halo}/M_{\rm sph}=100$ (right-hand panel). Note once again the steep
drop of all the curves around $\theta\simeq 1^\circ$ caused by the
absence of nearby sources of this kind and the very small contribution
given by the low-mass objects, mostly due to their small bias function
and their modest contribution to the effective volume emissivity.

In order to evaluate the total contribution of the clustering to
intensity fluctuations one has to add up all the values of $C(\theta)$
in eq.~({\ref{eq:cth}) obtained for the different mass intervals and
also to take into account the cross-correlation terms between objects
of different masses, so that the final expression for $C(\theta)$ is
given by
\begin{eqnarray}
C^{TOT}(\theta)=\sum_{i,j=1}^3 {\sqrt {C_i(\theta)C_j(\theta)}},
\label{eq:ctot}
\end{eqnarray}
where the indexes $i$,$j$ stand for high, intermediate and low masses.

\subsection{Power Spectrum of Temperature Fluctuations}
The angular power spectrum of the intensity fluctuations
[eqs. (\ref{eq:cth})--(\ref{eq:ctot})] can be obtained starting from
the well known expression for the expansion of temperature
fluctuations in spherical harmonics
\begin{eqnarray}
\frac{\delta T}{T}(\theta,\phi)=\sum_{l=0}^\infty\sum_{m=-l}^{l}a_l^m
Y_l^m(\theta,\phi).
\end{eqnarray}
If the fluctuations are a stationary process it can be shown (Peebles
1993) that the previous expression does not depend on the index $m$
(i.e. on the $\phi$ angle), so that we can write:
\begin{eqnarray}
C_l=\frac{1}{2l+1}\sum_{m=-l}^l\langle |a_l^m|^2\rangle=\langle
|a_l^0|^2 \rangle,
\end{eqnarray}
with
\begin{eqnarray}
a_l^0=\int_0^{2\pi}\int_0^\pi\frac{\delta
T(\theta)}{T}\sqrt{\frac{2l+1} {4\pi}}\;
P_l(\rm{cos}\theta)\;\rm{sin}\theta\;d\theta\;d\phi
\end{eqnarray}
(see Toffolatti et al. 1998), where $P_l({\rm cos}\theta)$ are the
Legendre polynomials, and $\delta T/T$ is expressed as
\begin{eqnarray}
\frac{\delta T}{T}(\theta)=\langle\left(\frac{\Delta T}{T}\right)^2
\rangle^{1/2} \!\!\!\!\!&=&\!\!\!\!\frac{\lambda^2\!
\sqrt{C^{TOT}(\theta)}}{2\;k_bT}\left[{\rm exp}\left (\frac{h\nu}{k_b
T}\right)-1 \right]^2\nonumber\\ &\!\!\!\!\!\!\!\times&
\!\!\!\!\!\!{\rm exp}\left(- \frac{h\nu}{k_b
T}\right)/\left(\frac{h\nu}{k_b T} \right)^2,
\end{eqnarray}
which relates intensity and temperature fluctuations (T=2.726 K,
Mather et al. 1994).

Figure \ref{fig:dT} shows the predicted values for the quantity
$\delta T_l=\sqrt{l(l+1)C_l/2\pi}$ (in units of K) at 353 GHz
($850\,\mu$m) - the central frequency of one of the channels of the
High Frequency Instrument (HFI) of the ESA's Planck mission - as a
function of the multipole $l$ up to $l=1000$. Results are plotted for
two values of the source detection limit ($S_d=100$ and 10~mJy; note
that the source detection limit for this Planck channel is likely to
be $>100\,$mJy) and the usual two values of $M_{\rm halo}/M_{\rm
sph}$.  Also shown, for comparison, is the power spectrum of primary
(CMB) anisotropies (solid line) predicted by a standard Cold Dark
Matter model for a $\Lambda$CDM ($\Lambda=0.7$, $\Omega_0=0.3$,
$h_0=0.7$), computed with the CMBFAST code developed by Seljak \&
Zaldarriaga (1996).

\begin{figure}
\vspace{8cm} \includegraphics{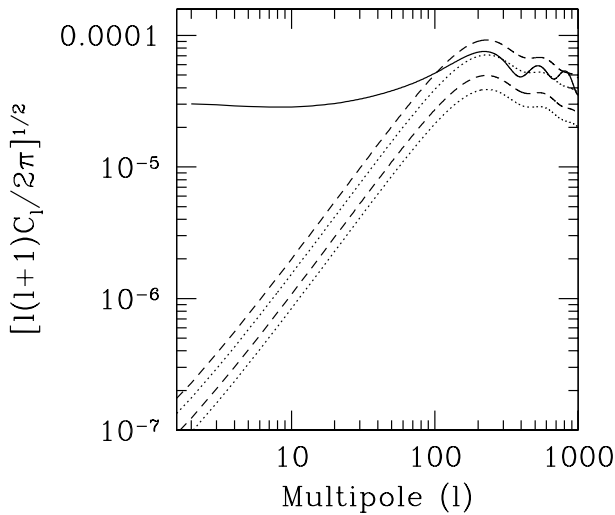}
\caption{Predicted power spectrum of temperature fluctuations $\delta
T_l=\sqrt{l(l+1)C_l/2\pi}$ (in units of K) as a function of the
multipole $l$ at 353 GHz, the central frequency of a Planck/HFI
channel.  The dashed lines are for a detection limit $S_d=100$~mJy,
the dotted ones for $S_d=10$~mJy. In both cases the higher curves are
obtained for $M_{\rm halo}/M_{\rm sph}$=100, the lower ones for
$M_{\rm halo}/M_{\rm sph}$=10. The solid line represents the power
spectrum of primary CMB anisotropies as predicted by a standard Cold
Dark Matter model for a $\Lambda$CDM cosmology ($\Lambda=0.7$,
$\Omega_0=0.3$, $h_0=0.7$).
\label{fig:dT}}
\end{figure}  

At this frequency our model predicts fluctuations of amplitude due to
clustering comparable to (and possibly even larger than) those
obtained for primary CMB anisotropies at $l\simgt 50$.  As illustrated
by Fig.~\ref{fig:lowb}, most of the clustering signal comes from
massive galaxies with fluxes $S\simgt 1$~mJy, which lie at substantial
redshifts (see Fig. \ref{fig:N_z}) and are therefore highly biased
tracers of the underlying mass distribution.  Also the strongly
negative K-correction increases their contribution to the effective
volume emissivity [eq.~(\ref{eq:jeff})] and therefore to the
fluctuations.\\ We therefore expect this effect to give a substantial
contribution to the power spectrum extracted e.g. from the BOOMERanG
map (de Bernardis et al. 2000) at 400 GHz. For $l \simgt 200$ its
contribution may be comparable to that due to inhomogeneous emission
by the interstellar dust.

\section{Conclusions}
We have worked out predictions for the clustering properties of
SCUBA-selected galaxies in the framework of an astrophysically
grounded model, relating the formation of QSOs and spheroids
(ellipticals, S0 galaxies and bulges of spirals). The theoretical
angular correlation function has been derived for different bias
functions, corresponding to different values of the ratio $M_{\rm
halo}/M_{\rm sph}$ between the mass of a spheroid locked in stars and
the mass of its host halo.

SCUBA sources are predicted to be strongly clustered, with a
clustering strength which increases with increasing mass (or
equivalently with increasing luminosity), as a consequence of the fact
that they are very massive and shining at substantial redshifts.
Since the clustering amplitude strongly depends on the quantity $M_{\rm
halo}/M_{\rm sph}$, future measurements of the angular correlation function 
$w(\theta)$ will be able to discriminate amongst different models of SCUBA 
galaxies and in particular to determine the amount of baryonic mass 
actively partaking the process of star formation. 

Under the hypothesis of low- to intermediate-mass primeval spheroids
to show up in the optical band as LBGs (as argued by Granato et
al. 2000), we have then compared our predictions with the clustering
measurements obtained by Giavalisco et al. (1998) for a wide sample of
LBGs at $z\simeq 3$. The agreement is good for a mass ratio between
the dark halo and the stellar component in the range 10--100; high
values for this ratio seem to be favoured.  \\ Also, the predicted
amplitude of the angular correlation function of SCUBA galaxies is
consistent with the one determined by Daddi et al.  (2000) for EROs
with $R-K_s \geq 5.3$ and $K_s \leq 19.2$. This would support the 
observational evidence for a substantial fraction of SCUBA galaxies to be 
identified with EROs (Smail et al. 1999; Ivison et al. 2000).

We have also considered the effect of clustering on maps of the Cosmic
Microwave Background at sub-mm wavelengths, with special attention to
the 353~GHz ($850\,\mu$m) channel of the High Frequency Instrument
(HFI) of the ESA's Planck mission. The possibility of large
fluctuations due to clustering at this (and also shorter) wavelength
was first pointed out by Scott \& White (1999) who adopted a
phenomenological approach, assuming for SCUBA galaxies the clustering
amplitudes observed for LBGs (Giavalisco et al. 1996). Our treatment,
leading to a derivation of the angular autocorrelation function in the
framework of the currently standard hierarchical clustering scenario,
indicates that massive SCUBA galaxies (expected to dominate the population 
of sources sampled at fluxes $S\simgt 10$~mJy) are likely to be even more 
strongly clustered than LBGs. The corresponding fluctuations are then expected
to be, at $\simeq 350\,$GHz, of an amplitude comparable with those due
to primary CMB anisotropies for multipole numbers $l > 50$. This would
imply that important information on the clustering properties of faint
sub-mm galaxies (and hence on their physical properties such as their
mass and/or the amount of baryons involved in the star-formation
process) resides in the BOOMERanG maps at 400 GHz where, however, the
dominant signal is expected to come from interstellar dust
emission. On the other hand, if the ratio $M_{\rm halo}/M_{\rm sph}$
is large, we expect the clustering signal to be comparable with dust
emission fluctuations at $l \simgt 200$. It might therefore be
possible to separate these two contributions and possibly exploit the
different shapes of their power spectra.

As stressed by Scott \& White (1999), the amplitude of the clustering
signal drops rapidly - in comparison to the one arising from primary
anisotropies - with decreasing frequency. In the nearest Planck/HFI
channel (centered at 217 GHz) its contamination on the CMB anisotropy
maps should already be negligible.

\noindent
\section*{ACKNOWLEDGMENTS}   
Mauro Giavalisco is warmly thanked for providing us with the data shown in 
Figure 3. We are grateful to Pierluigi Monaco for useful clarifications on the 
relationship between SCUBA galaxies and quasars. 
We also thank ASI and Italian MURST for financial support.

\end{document}